\documentclass[sigconf]{acmart}

\usepackage{booktabs} 
\usepackage{subfigure}
\usepackage{listings}
\usepackage{hyperref}

\acmConference[Second International Symposium on Checkpointing for Supercomputing]{SC'21: SuperCheck}{November 16--18, 2021}{St. Louis, MO}
{\description \item[\fbox{To Do}]~\sl%
  \typeout{----------- Something left to be done here!
    ------------------}}%
{\enddescription}%

\begin{document}
\title{Towards Aggregated Asynchronous Checkpointing}

\author{Mikaila J. Gossman}
\affiliation{%
  \institution{Clemson University}
  \country{USA}
}
\email{mikailg@g.clemson.edu}

\author{Bogdan Nicolae}
\affiliation{%
  \institution{Argonne National Laboratory}
  \country{USA}
}
\email{bnicolae@anl.gov}

\author{Jon C. Calhoun}
\affiliation{%
  \institution{Clemson University }
  \country{USA}
}
\email{jonccal@clemson.edu}

\author{Franck Cappello}
\affiliation{%
  \institution{Argonne National Laboratory}
  \country{USA}
}
\email{cappello@anl.gov}

\author{Melissa Smith}
\affiliation{%
  \institution{Clemson University }
  \country{USA}
}
\email{smithmc@clemson.edu}

\renewcommand{\shortauthors}{M. Gossman et al.}

\begin{abstract}
     High-Performance Computing (HPC) applications need to checkpoint massive amounts of data at scale. Multi-level asynchronous checkpoint runtimes like VELOC (Very Low Overhead Checkpoint Strategy) are gaining popularity among application scientists for their ability to leverage fast node-local storage and flush independently to stable, external storage (e.g., parallel file systems) in the background. Currently, VELOC adopts a one-file-per-process flush strategy,
     which results in a large number of files being written to external storage, thereby
     overwhelming metadata servers and making it difficult to transfer and access checkpoints as a whole. This paper discusses the viability and challenges of designing aggregation techniques for asynchronous multi-level checkpointing. To this end we implement and study two aggregation strategies, their limitations,
     and propose a new aggregation strategy specifically for asynchronous multi-level
     checkpointing.
\end{abstract}


\maketitle

\section{Introduction}
\label{sec:intro}

Checkpointing distributed HPC applications is a common task in various scenarios: resilience, job management, reuse of computational 
states. Typically, checkpoints are persisted to an external repository such as a parallel file system (PFS), whose aggregated I/O bandwidth is limited. Synchronous checkpointing strategies that block applications until all processes checkpoint to a PFS are projected to be impractical at exascale due to high overheads~\cite{veloc}. To alleviate this issue, asynchronous multi-level checkpoint systems like VELOC~\cite{veloc} write checkpoints to fast, node-local storage and flushes them to the PFS in the background while the application resumes, thereby reducing the checkpoint overhead.

While flushing to the PFS, asynchronous checkpointing schemes share resources of the compute nodes (CPU cores, memory, network bandwidth) with the application processes, creating contention. Therefore, an important challenge is to maintain high throughput to the PFS in the presence of resource sharing, and mitigate the resource contention to avoid additional overheads that slow the application or extends the flush time. In this regard, VELOC adopts a one-file-per-process asynchronous flush strategy, which has two advantages: (1) it is simple, portable and efficient (no coordination, synchronization, or I/O locks); (2) it enables an independent strategy for mitigating contention on each compute node.

However, file-per-process strategies have limitations due to generating 
large numbers of files at scale and overwhelming the PFS. Specifically, a PFS
experiences metadata bottlenecks when many files are accessed simultaneously (especially when stored in the same directory)~\cite{6005432}. Furthermore, it is difficult for developers to manage massive amounts of checkpoint files,
especially in scenarios that involve: moving between data centers, verification 
of integrity, use in a producer-consumer workflow, etc. To alleviate these limitations, 
checkpoint aggregation (e.g. GenericIO~\cite{Habib_2016}) can be used to combine checkpoints from $N$ processes into $M$ files, with $M$ typically set by the 
application and often fixed to one.

Although effective for synchronous checkpointing~\cite{MCREngine, AsyncInterferenceJHPCA-21, Habib_2016}, such aggregation strategies, to our knowledge, remain unexplored for asynchronous checkpointing. The main challenge in this context is designing efficient
aggregation strategies that achieve high I/O bandwidth under concurrency 
to maintain a high speed of flushing, while also minimizing overhead caused by resource contention. This is a non-trivial challenge that involves a co-design of both aggregation and mitigation strategies. 

In this paper, we discuss the benefits and challenges of solving the co-design
problem mentioned above. Specifically, we focus on several aggregation strategies for
VELOC~\cite{veloc_abs,veloc}, a production-ready checkpointing system used on
large-scale HPC systems. Using these strategies, we identify key bottlenecks in 
the checkpointing process (Section~\ref{sec:strategies}) and discuss several future work
directions to address them (Section~\ref{sec:future}).

\section{Aggregation Strategies}
\label{sec:strategies}

In a study of asynchronous I/O interference, Tseng et. al.~\cite{AsyncInterferenceJHPCA-21} found that there is a trade-off between
increasing the number of concurrent I/O threads (to flush to external storage faster) vs. 
the slowdown perceived by the application. Furthermore, there are trade-offs 
between the OS-level overheads incurred by the flushing strategy (read/write, 
mmap/write, sendfile) and the underlying PFS I/O implementation. For example,
even if sendfile triggers the least context switches and therefore minimizes
OS-level overheads, it performs small I/O requests to the PFS, which are
handled by the latter inefficiently, thereby increasing both the application 
slowdown and the flush duration. These trade-offs are important to consider in
the design of asynchronous checkpoint aggregation strategies. 

\subsection{POSIX-based aggregation}
\label{sec:strategy:posix}
The first strategy we implement is a POSIX-based aggregation strategy that
aims to highlight the bottlenecks during asynchronous flushing. To this end,
we leverage the \emph{multi-level} checkpointing support introduced by 
VELOC~\cite{veloc_abs}: (1) each process writes its checkpoint in a blocking 
fashion to a file on node-local storage, e.g. in-memory or to an SSD (denoted the \emph{local phase}); (2) once files are written locally, the process continues with application computation and 
a separate asynchronous post-processing engine (denoted the \emph{active backend}
in VELOC) running on each compute node 
writes the node-local files (generated by processes co-located on the same
node) to an offset in a shared file located on the PFS (denoted the \emph{flush phase}). 
The offset of each
node-local checkpoint in the remote file is calculated using a prefix-sum
algorithm over all participating processes. Note that with this strategy, the active backend 
running on each node can parallelize writing the local checkpoints using a number of 
I/O threads that can be used to match
the desired trade-off between application slowdown and flushing duration. 
This can be configured for each active backend independently.

Thanks to its simplicity, the POSIX-based aggregation strategy incurs minimal
preparation and synchronization overheads, much like VELOC's default
asynchronous strategy that writes each checkpoint to a different file on
the PFS (i.e. no intercommunication between processes during the flush phase). Therefore, we can conclude that this strategy is unlikely to cause more interference or slowdown in the presence of an application. However, the flushing duration is significantly higher compared to the default VELOC and GenericIO. This extended flush time is caused by \emph{false sharing}: to facilitate parallel processing, files
are striped on the PFS into chunks that are stored on different I/O
servers. If the writes of local checkpoints do not align to the stripe sizes, then
the writes will compete for the same stripe, despite using non-overlapping file regions.
As a consequence, more advanced POSIX-based aggregation strategies need to address
the negative impact of false sharing.

\subsection{MPI-IO based aggregation}
\label{sec:strategy:mpi-io}
A second strategy we implement on top of VELOC is based on MPI-IO, which is also used by GenericIO
and was optimized for synchronous aggregation. Specifically, the same prefix-sum algorithm is applied to determine the offset of each node-local checkpoint 
in the larger file on the PFS, then the active backend issues a collective MPI-IO \emph{write at} operation (instead 
of independent POSIX writes). The underlying collective write implementation is
based on two key observations: (1) the number of processes (MPI ranks) is typically
much larger than the number of I/O servers, therefore it is suboptimal to issue
more concurrent I/O writes than there are I/O servers available to execute them;
(2) only one MPI rank is allowed to write to a PFS stripe due to false sharing. Based on these observations, MPI-IO implements optimizations
such as gathering the data from multiple MPI ranks on \emph{I/O leaders},
which are then responsible for concurrently writing files on the PFS. The number
of I/O leaders can be adjusted to match the number of I/O servers. Furthermore, 
each leader can be assigned a set of stripes that is disjoint from the 
sets of stripes of all other leaders, thereby eliminating false sharing.

While more advanced, an MPI-IO based aggregation strategy has limited applicability
in the area of asynchronous checkpointing for several reasons. First, it introduces
high communication overheads which is needed to gather the data from multiple processes to the 
I/O leaders. While this does not cause any interference in synchronous mode (because the application is 
blocked during checkpointing), it can become a bottleneck in asynchronous mode,
because the application may need to perform its own communications during the
gathering phase, thereby competing for network bandwidth. Second, since it is 
based on a collective operation, it assumes all MPI ranks need to participate
in the checkpointing process and all data needs to be ready \emph{simultaneously}
in all MPI ranks, introducing synchronization during the flush phase. However, a key feature of multi-level checkpointing strategies is that the MPI ranks write their checkpoints to node-local storage and flush to the PFS independent of the progress or state of other participating processes,
then continue executing the application code. 

Thus, it is up to
the active backends to self-organize in order to optimize the aggregation
asynchronously. One potential solution is to perform a collective MPI-IO write
on all active backends, where each active backend contributes with its set of
node-local checkpoints. However, such a strategy introduces two bottlenecks:
(1) the MPI-IO standard does not allow more than a single contiguous 
memory region to be written to a file, whereas each active backend may need to
contribute with multiple such regions (mapped to independent node-local
checkpoint files); (2) even if such a MPI-IO collective write were implemented,
it places an unnecessary restriction on the active backends to be ready
at the same time to facilitate communication and data exchange, which increases 
the initialization overheads. To address these issues, we propose a multi-phase
solution that invokes multiple successive MPI-IO collective write calls, one for each
node-local checkpoint.

\subsection{Preliminary evaluation}
\label{sec:strategy:eval}

We evaluate the two strategies introduced in Section~\ref{sec:strategy:posix} and
Section~\ref{sec:strategy:mpi-io} in a series of experiments that involve
a micro-benchmark running on an increasing number of MPI ranks and an increasing
number of compute nodes. Each MPI rank writes a checkpoint that is 1~GiB large.
The testbed used for our experiments is Argonne National Lab's Theta supercomputer, 
a Cray XC40 11.60 Petaflops system, equipped with a Lustre parallel file system
~\cite{veloc_abs, veloc}. We compare the two strategies with two additional baseline
approaches: the default VELOC multi-level asynchronous strategy (one file per MPI rank)
and a synchronous I/O aggregation strategy using GenericIO (GIO)~\cite{Habib_2016}. The
aggregation performed by all approaches is $N \to 1$).

\begin{figure}
    \centering
    \includegraphics[width=0.5\textwidth]{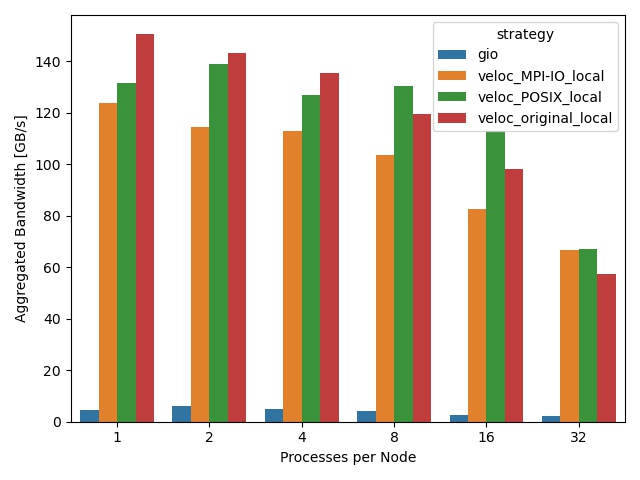}
    \caption{Local checkpointing phase (blocking) for an increasing number of processes
    per node. Note that GIO is writing directly to the PFS. Higher is better.}
    \label{fig:local_poster_results}
\end{figure}

\begin{figure}
    \centering
    \includegraphics[width=0.5\textwidth]{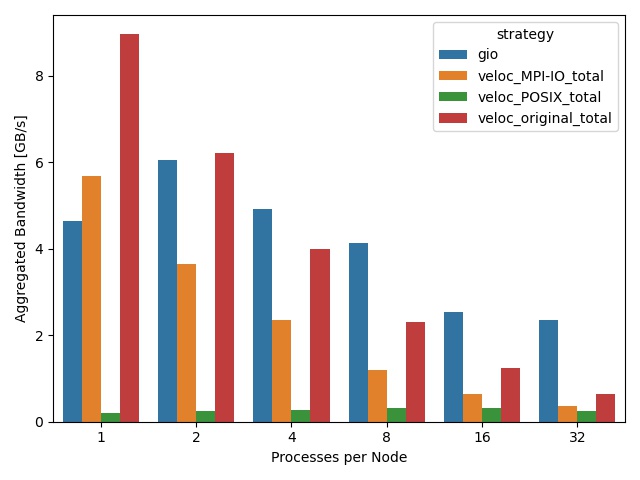}
    \caption{Flush phase to the PFS (async) for an increasing number of processes per node. 
    In the case of GIO, this is blocking and overlaps with the counterpart from Figure~ \ref{fig:local_poster_results}. Higher is better.}
    \label{fig:total_poster_results}
\end{figure}

Figure ~\ref{fig:local_poster_results} shows the throughput during the local checkpoint phase for the aforementioned checkpoint strategies. Since VELOC is writing to local storage (SSD or memory
hierarchy), it is orders of magnitudes faster than GenericIO, which is writing directly to the
PFS. Note that all three VELOC strategies exhibit similar throughput, which means the prefix
sum employed by the aggregation approaches introduces a negligible overhead during the local 
checkpointing phase. 

Figure \ref{fig:total_poster_results} shows the throughput for the compared strategies during the flush phase. As can be observed, both the POSIX and MPI-IO based strategies show a significantly
lower throughput compared with the original one file per MPI rank strategy, which can be 
attributed to false sharing in the case of the former, and, respectively, to the fact
that we invoke multiple MPI-IO collective calls, which perform suboptimally.

Thus, we can conclude that a simple aggregation strategy based on POSIX or MPI-IO is not
sufficient and there is significant room for improvement to reach and potentially even surpass
the default, embarrassingly parallel one-file per MPI rank flush strategy.

\section{Future work}
\label{sec:future}
Based on the study described in Section~\ref{sec:strategies}, we propose
an aggregation strategy specifically designed for asynchronous multi-level 
checkpointing. Assuming the checkpoints need to be flushed to a set of
$M$ fixed-sized stripes on the PFS (which may form a single or up to $M$ files),
we elect $M$ leaders among the active backends. Similar to synchronous 
aggregation strategies, the active backends that are not designated as leaders
will send their checkpointing data to one (or multiple) of the leaders. The decision of 
which active backends becomes a leader and who is supposed to send their 
checkpointing data can be determined dynamically based on several factors,
such as:
(1) size of the node-local checkpoints (e.g., an active backend with
small node-local checkpoints should send them to an active backend with
larger node-local checkpoints to minimize communication over the network); (2) the current
load of each compute node (e.g., nodes with the least CPU/memory/network
pressure are less likely to cause bottlenecks and therefore should become
leaders); (3) network topology (e.g., leaders should gather the 
checkpointing data from active backends that are in close proximity).

These decisions can be implemented by piggy-backing additional information during the prefix-sum. In the previously mentioned strategies, the prefix-sum operation calculates the offset in the shared, remote file. In this strategy, we will statically assign each leader a set of offsets aligned to the stripe size. The prefix-sum propagates through all the processes and informs non-leaders 1) who the leaders are; 2) how much data to send to each (since some non-leaders may contain more data than can fit into a single leader's remaining stripes, it may need to be broken up among various leaders).
Then, by applying the same strategy independently, the nodes can take different roles without the need for further agreement protocols, which has two advantages: (1) it is a lightweight protocol that
introduces a minimal runtime overhead; (2) this only requires active backends to synchronize to perform the leader election, alleviating one of the limitations of collective operations.

\section{Conclusion}
\label{sec:conclusions}
This paper discusses the opportunities for aggregation in asynchronous multi-level checkpointing and the challenges of designing efficient techniques in this context. To this end, we show that false
sharing exhibited by POSIX-based I/O writes to shared files on the PFS, as well as state-of-art optimization techniques for synchronous I/O aggregation (e.g. MPI-IO) suffer from bottlenecks and limitations when directly applied in the context of multi-level asynchronous checkpointing. As a consequence, we contribute a study of these limitations and outline the basic ideas behind a proposal
for a dedicated aggregation strategy specifically designed for asynchronous multi-level checkpointing that alleviates the identified limitations.

\begin{acks}
Clemson University and Argonne National Lab are acknowledged for generous allotment of compute time on the Palmetto cluster and Theta super computer. This material is based upon work supported by the National Science Foundation under Grant No. SHF-1910197. The material was supported by U.S. Department of Energy, Office of Science, under contract DE-AC02-06CH11.357
\end{acks}


\bibliographystyle{ACM-Reference-Format}
\bibliography{refs.bib}

\end{document}